\documentstyle[12pt,epsfig]{article}
\newlength{\dinwidth}
\newlength{\dinmargin}
\newcommand{\ba}{\begin{array}}
\newcommand{\ea}{\end{array}}
\newcommand{\beq}{\begin{equation}}
\newcommand{\eeq}{\end{equation}}
\newcommand{\bea}{\begin{eqnarray}}
\newcommand{\eea}{\end{eqnarray}}

\def\ep{\varepsilon}

\def\S{{\bf S}}

\def\bce{\begin{center}}
\def\ece{\end{center}}

\def\nonu{\nonumber}

\def\pa{\partial}

\def\be{\beta}
\def\ga{\gamma}

\def\de{\delta}
\def\De{\Delta}
\def\ep{\epsilon}

\def\la{\lambda}
\def\La{\Lambda}

\def\S{{\bf S}}

\begin{document}
\thispagestyle{empty}
\addtocounter{page}{-1}
\begin{flushright}
{\tt hep-th/0008065}\\
\end{flushright}
\vspace*{1.3cm}
\centerline{\Large \bf Three-Dimensional SCFTs, Supersymmetric 
Domain Wall  }
\vskip0.3cm
\centerline{\Large \bf and Renormalization Group Flow}
\vspace*{1.5cm} 
\centerline{\bf Changhyun Ahn {\rm and} Jinsub Paeng}
\vspace*{1.0cm}
\centerline{\it Department of Physics, 
Kyungpook National University, Taegu 702-701 Korea}
\vskip0.3cm
\vspace*{0.8cm}
\centerline{\tt ahn@knu.ac.kr}
\vskip2cm
\centerline{\bf abstract}
\vspace*{0.5cm}

By analyzing $SU(3) \times U(1)$ invariant stationary point, studied earlier
by Nicolai and Warner, of gauged
${\cal N}=8$ supergravity, we find that the deformation of $\S^7$ gives rise to
nontrivial 
renormalization group flow in a three-dimensional boundary
super conformal field theory from ${\cal N}=8$, $SO(8)$ invariant UV fixed 
point to ${\cal N}=2$, $SU(3) \times U(1)$ invariant IR fixed point. 
By explicitly 
constructing 28-beins $u, v$ fields, that are an element of fundamental 
56-dimensional representation of $E_7$, in terms of scalar and pseudo-scalar
fields of gauged ${\cal N}=8$ supergravity, we get $A_1, A_2$ tensors. 
Then we identify one of the eigenvalues of $A_1$ tensor 
with ``superpotential'' of
de Wit-Nicolai scalar potential and discuss four-dimensional
supergravity description of renormalization group flow,
i.e.  the BPS domain wall solutions 
which are equivalent to vanishing of variation of spin 1/2, 3/2 fields
in the supersymmetry preserving bosonic background 
of gauged ${\cal N}=8$ supergravity.
A numerical analysis of the steepest descent equations interpolating two 
critical points is given.


\baselineskip=18pt
\newpage

\section{Introduction}
\setcounter{equation}{0}
Few examples are known for three-dimensional interacting conformal field 
theories, mainly due to strong coupling dynamics in the infrared(IR) limit. 
In the previous papers \cite{ar,ar1},  
three-dimensional (super)conformal field theories were classified by utilizing 
the AdS/CFT correspondence \cite{maldacena,witten,gkp} and earlier, exhaustive 
study of the Kaluza-Klein supergravity \cite{duff}.

The simplest spontaneous compactification of the eleven-dimensional 
supergravity \cite{cremmer} is the Freund-Rubin \cite{fr} compactification
to a product 
of $AdS_4$ space-time and an arbitrary compact Einstein 
manifold $X_7$ of positive scalar curvature. 
The best known example is provided by round- and squashed-$\S^7$. 
The standard 
Einstein metric of the round-$\S^7$ yields a vacuum with $SO(8)$ gauge 
symmetry and ${\cal N}=8$ supersymmetry. The second, 
squashed Einstein metric \cite{jensen}, yields a vacuum with $SO(5) \times 
SO(3)$ gauge symmetry and ${\cal N}=1$( or 0) supersymmetry, 
depending on the orientation of the $\S^7$ \cite{awada}.
In \cite{ar}, the well-known spontaneous (super)symmetry 
breaking deformation from round- to squashed-$\S^7$ was mapped to a 
renormalization group(RG) flow from ${\cal N}=0$( or 1), $SO(5) \times SO(3)$ 
invariant fixed point in the ultraviolet(UV) 
to ${\cal N}=8$, $SO(8)$ invariant fixed 
point in the IR. The squashing 
deformation corresponded to an irrelevant operator at the ${\cal N}=8$ 
superconformal fixed point and a relevant operator at the ${\cal N}=1$( 
or 0) (super)conformal fixed point, respectively.

In contrast to the Freund-Rubin compactifications, the symmetry 
of the vacuum of Englert type compactification is no longer given by
the isometry group of $X_7$ but rather by the group which leaves invariant
both the metric and four-form magnetic field strength. By 
generalizing compactification vacuum ansatz to the nonlinear level,
solutions of the eleven-dimensional supergravity were obtained 
directly from the scalar and pseudo-scalar expectation values at
various critical points of the ${\cal N}=8$ supergravity potential \cite{dnw}. 
They reproduced all known Kaluza-Klein solutions of 
the eleven-dimensional supergravity: round $\S^7$ \cite{dp}, 
$SO(7)^-$-invariant, {\sl parallelized} $\S^7$ \cite{englert},
$SO(7)^{+}$-invariant vacuum \cite{dn}, $SU(4)^{-}$-invariant vacuum 
\cite{pw}, and a new one with $G_2$ invariance. 
Among them, round $\S^7$- and $G_2$-invariant vacua are
stable, while $SO(7)^{\pm}$-invariant ones are known to be unstable 
\cite{dn1}. In \cite{ar1}, via AdS/CFT correspondence, deformation of 
$\S^7$ was interpreted as renormalization group flow from ${\cal N}=8$,
$SO(8)$ invariant UV fixed point to ${\cal N}=1$, $G_2$ invariant
IR fixed point by analyzing de Wit-Nicolai potential.

In this paper, we will continue to analyze
a vacuum of ${\cal N}=8$ supergravity with $SU(3) \times U(1)$ symmetry, 
studied earlier by Nicolai and Warner \cite{nw},
that was considered very briefly in \cite{ar1}.
In section 2, by explicitly constructing 28-beins $u, v$ fields, 
that are an element of fundamental 
56-dimensional representation of $E_7$, in terms of scalar and pseudo-scalar
fields of ${\cal N}=8$ supergravity, we get $A_1, A_2$ tensors 
(\ref{a1}), (\ref{yis}). 
In section 3, 
we will be identifying a deformation which gives 
rise to a renormalization group flow associated with the symmetry breaking 
$SO(8) \rightarrow SU(3) \times U(1)$(both of which are stable vacua) 
and find that the deformation operator is relevant at the $SO(8)$ 
fixed point but becomes irrelevant at the $SU(3) \times U(1)$ fixed point.
In section 4,
we identify one of the eigenvalues of $A_1$ tensor with ``superpotential'' of
de Wit-Nicolai scalar potential and discuss the BPS domain wall solutions.
Finally in appendix, there exist some details.  
See also recent papers \cite{ads5} on RG flows and AdS/CFT correspondence.

\section{ de Wit-Nicolai  Potential and 
${\rm AdS}_4$ Supergravity Vacua}

de Wit and Nicolai \cite{dn3,dn2} 
constructed a four-dimensional supergravity theory
by gauging the $SO(8)$ subgroup of $E_7$ in the global $E_7$ $\times$
local $SU(8)$ supergravity of Cremmer and Julia \cite{cremmer}. In common
with Cremmer-Julia theory, this theory contains self-interaction of a single
massless ${\cal N}=8$ supermultiplet of spins 
$(2, 3/2, 1, 1/2, 0^{+}, 0^{-})$ {\sl but} with local $SO(8)$ $\times$ local
$SU(8)$ invariance. 
It is well known \cite{cj} that the 70 real 
scalars of ${\cal N}=8$ supergravity
live on the coset space $E_7/SU(8)$ since 63 fields may be gauged
away by an $SU(8)$ rotation and are described by 
an element ${\cal V}(x)$ of the fundamental 56-dimensional representation
of $E_7$:
\bea
{\cal V}(x)=
\left(
\begin{array}{cc}
u_{ij}^{\;\;IJ}(x) & v_{ijKL}(x)  \\
v^{klIJ}(x) & u^{kl}_{\;\;KL}(x)
\end{array} 
\right), 
\label{56bein}
\eea
where $SU(8)$ index pairs $[ij], \cdots$ and $SO(8)$ index pairs $[IJ], \cdots$
are antisymmetrized and therefore $u$ and $v$ fields
are $28 \times 28$ matrices and $x$ is the coordinate on 4-dimensional 
space-time.
Complex conjugation can be done by raising or lowering those
indices, for example,
$(u_{ij}^{\;\;IJ})^{\star}= u^{ij}_{\;\;IJ}$ and so on.
Under local $SU(8)$ and local $SO(8)$, the matrix ${\cal V}(x)$ transforms
as ${\cal V}(x) \rightarrow U(x) {\cal V}(x) O^{-1}(x)$ where 
$U(x) \in SU(8)$ and $O(x) \in SO(8)$ and matrices $U(x)$ and 
$O(x)$ are in the appropriate 56-dimensional representation.
In the gauged supergravity theory, the 28-vectors transform in the
adjoint of $SO(8)$ with resulting non-abelian field strength.

Although the full gauged ${\cal N}=8$ Lagrangian is rather complicated 
\cite{dn2},
the scalar and gravity part of the action  is simple(we are considering
a gravity coupled to scalar field theory since matter fields do not
play a role in domain wall solutions of section 4)
and maybe written as 
\bea
\int d^4 x \sqrt{-g} \left( \frac{1}{2} R 
- \frac{1}{96} \left| A_{\mu}^{\;\;ijkl} \right|^2 - 
V \right)
\label{action}
\eea
where the scalar kinetic terms are completely antisymmetric and self-dual
in its indices:
\bea
A_{\mu}^{\;\; ijkl} = -2 \sqrt{2} \left( u^{ij}_{\;\;IJ} 
\partial_{\mu} v^{klIJ} -
v^{ijIJ} \partial_{\mu} u^{kl}_{\;\;IJ} \right)
\label{aijkl}
\eea
and $\left| A_{\mu}^{\;\;ijkl} \right|^2$ is a product of 
$A_{\mu}^{\;\;ijkl}  $ and its complex conjugation
as above and $\mu$ is the 4-dimensional space-time index.
Let us define $SU(8)$ so called T-tensor which is cubic in the 
28-beins $u$ and $v$, manifestly antisymmetric in the indices
$[ij]$ and $SU(8)$ covariant. This comes from naturally by introducing
a local gauge coupling in the theory. Furthermore, other tensors coming from
T-tensor play an important role in this paper and scalar structure is
encoded in two $SU(8)$ tensors. That is, 
$A_1^{\;\;ij}$ tensor is symmetric in $(ij)$
and $A_{2l}^{\;\;\;ijk}$ tensor is antisymmetric in $[ijk]$: 
\bea
T_l^{\;kij} & = & 
\left(u^{ij}_{\;\;IJ} +v^{ijIJ} \right) \left( u_{lm}^{\;\;\;JK} 
u^{km}_{\;\;\;KI}-v_{lmJK} v^{kmKL} \right), \nonu \\
A_1^{\;\;ij} & =& 
-\frac{4}{21} T_{m}^{\;\;ijm}, \;\;\; A_{2l}^{\;\;\;ijk}=-\frac{4}{3}
T_{l}^{\;[ijk]}.
\label{ttensor}
\eea

Then de Wit-Nicolai effective nontrivial
potential arising from $SO(8)$ gauging can be written as compact form:
\bea
V= -g^2 \left( \frac{3}{4} \left| A_1^{\;ij} \right|^2-\frac{1}{24} \left|
A^{\;\;i}_{2\;\;jkl}\right|^2 \right) 
\label{V}
\eea
where
$g$ is a $SO(8)$ gauge coupling constant and it is understood that
the squares of absolute values of $A_1, A_2$ are nothing but a product of
those and its complex conjugation on 28-beins $u$ and $v$.
The 56-bein ${\cal V}(x)$ can be brought into the following form by the 
gauge freedom of $SU(8)$ rotation
\bea
{\cal V}(x)=
\mbox{exp} \left(
\begin{array}{cc}
0 &  \phi_{ijkl}(x)  \\
 \phi^{mnpq}(x) & 0 
\end{array} \right),
\label{calV}
\eea  
where $\phi^{ijkl}$ is a complex self-dual tensor describing the 35 
scalars $\bf 35_{v}$(the real part of $\phi^{ijkl}$)
and 35 pseudo-scalar fields $\bf 35_{c}$(the imaginary part of $\phi^{ijkl}$)
  of ${\cal N}=8$ supergravity.
After gauge fixing, one does not distinguish between $SO(8)$ and $SU(8)$
indices. 
The scalar potential of gauged ${\cal N}=8$ 
supergravity has four stationary
points with at least $G_2$ invariance \cite{warner}.
The full supersymmetric solution where both $\bf 35_{v} $ 
scalars and $\bf 35_{c} $ pseudo-scalars
vanish yields $SO(8)$ vacuum state wih ${\cal N}=8$ supersymmetry(Note 
that $SU(8)$ is {\it not} a symmetry of the vacuum). 

It is known that, in ${\cal N}=8$ supergravity, there also exists 
a ${\cal N}=2$ supersymmetric, $SU(3) \times U(1)$ invariant vacuum \cite{nw}. 
To reach this critical point, one has to turn on expectation values of both 
scalar $\la(x) $ and pseudo-scalar $\la'(x)$ fields as
\bea
\langle \phi_{ijkl}(x) \rangle =
\frac{1}{2 \sqrt{2}} \left( \la(x) \; X^{+}_{ijkl}
+i \la'(x) \; X^{-}_{ijkl} \right), \nonu
\eea 
where
$ X^{+}_{ijkl}$ and $ X^{-}_{ijkl} $ are the 
unique(up to scaling), completely antisymmetric self-dual and anti-self-dual
tensors which are invariant under $SU(3) \times U(1),$
\bea
  X^{+}_{ijkl} &=& +[ (\de^{1234}_{ijkl}+\de^{5678}_{ijkl})+
 (\de^{1256}_{ijkl}+ \de^{3478}_{ijkl})+(\de^{1278}_{ijkl}
 +\de^{3456}_{ijkl})]
\nonu \\
       X^{-}_{ijkl} &=& -[(\de^{1357}_{ijkl}
-\de^{2468}_{ijkl})+(\de^{1368}_{ijkl}
     -\de^{2457}_{ijkl})+(\de^{1458}_{ijkl} -\de^{2367}_{ijkl})-
   (\de^{1467}_{ijkl}-\de^{2358}_{ijkl})]. \nonu
\eea
Therefore 56-beins ${\cal V}(x)$ can be written as $ 56\times 56$ matrix whose
elements are some functions of scalar and pseudo-scalars by exponentiating
the vacuum expectation value $\phi_{ijkl}$. On the other hand, 28-beins $u$
and $v$ are an element of this ${\cal V}(x)$ according to (\ref{56bein}).
One can construct 28-beins $u$ and $v$ in terms of $\la(x)$ and 
$\la'(x)$ fields
explicitly and they are given in the appendix (\ref{uv}). 
Now it is ready to get the informations on the $A_1$ and $A_2$ tensors
in terms of $\la(x)$ and $\la'(x)$ via (\ref{ttensor}). 

It turned out that
$A_1^{\;IJ}$ tensor has two distinct eigenvalues, $z_1$, $z_2$ with
degeneracies 6, 2 respectively and has the following form
\bea
A_1^{\;IJ} =\mbox{diag} \left(z_1, z_1, z_1, z_1, z_1, z_1, z_2,
z_2 \right), 
\label{a1}
\eea
where
\bea
z_1  =  \frac{1}{4} (p+q) \left( c(3+m')-2s \right), \qquad
z_2  =  \frac{1}{4} (p+q) \left( 
c(3+m')-2s m' \right) \nonu
\eea
and
\bea
& & 
p \equiv \cosh(\la/2\sqrt{2}), \;\; q \equiv \sinh(\la/2\sqrt{2}), \;\;
p' \equiv \cosh(\la'/2\sqrt{2}), \;\; q' \equiv \sinh(\la'/2\sqrt{2}) \nonu \\
& & m \equiv \cosh(\sqrt{2} \la), \;\; n \equiv \sinh(\sqrt{2} \la),\;\;
m' \equiv \cosh(\sqrt{2} \la'), \;\; n' \equiv \sinh(\sqrt{2} \la') \nonu \\
& & c \equiv \cosh(\la/\sqrt{2}), \;\; s \equiv \sinh(\la/\sqrt{2}), \;\;
c' \equiv \cosh(\la'/\sqrt{2}), \;\; s' \equiv \sinh(\la'/\sqrt{2}). 
\label{sinhcosh}
\eea
The eigenvalue $z_2$ at the $SU(3) \times U(1)$ critical point is equal to
$-\sqrt{- \frac{\La_{SU(3) \times U(1)}}{6}} 
\frac{\ell_{\rm pl}}{g}$ where 
$\La_{SU(3) \times U(1)}$ is the cosmological constant at that point,
as we will see later. So it has unbroken ${\cal N}=2$ supersymmetry:the
number of supersymmetries is equal to the number of eigenvalue of $A_1$
tensor for which $|\mbox{eigenvalue of} A_1 |=\ell_{\rm pl}\sqrt{-\frac{\La}
{6 g^2 }}$. On the other hand, at the $SO(8)$ critical point, $z_1=z_2$ and
since these 8 eigenvalues are equal to $-\ell_{\rm pl}\sqrt{-\frac{\La}
{6 g^2 }}$, this gives ${\cal N}=8$ supersymmetry as we expected.

Similarly $A_{2, L}^{\;\;\;\;\;IJK}$ 
tensor can be obtained from the triple product of
$u$ and $v$ fields by definition (\ref{ttensor}). 
It turns out that they are written in terms of
five kinds of 
fields
$y_1, y_2, y_3, y_4$ and $y_5$ and are given in the appendix:
\bea
& & y_1  =  \frac{1}{4} ( p+q) \left( c(1-m')-2s \right), \;\;\;
y_2  =  \frac{1}{4} ( p+q) \left( c(1-m')+2s \right)=\frac{2\sqrt{2}}{3}
\frac{\partial z_2}{\partial \la}, \nonu \\
& & y_3  =  \frac{1}{4} ( p+q) \left( c-m'(c-2s) \right),\;\;\;
y_4  =  -\frac{1}{4} i ( p+q) c n', \nonu \\
& & y_5  =  -\frac{1}{4} i (p+q) (c-2s)n'= \frac{-i}{\sqrt{2}} 
\frac{\partial z_2}{
\partial \la'}. 
\label{yis} 
\eea
Note the expression of $z_2$ and its derivatives with respect to $\la$ and
$\la'$ which are $y_2$ and $y_5$ up to some constant, 
respectively. 
This observation will be crucial as we discuss about supersymmetry variations
in the context of BPS domain wall solutions in section 4. 
One of the eigenvalues of
$A_1^{\;\;IJ}$ tensor, $z_2$ will provide a ``superpotential'' of $V$ in 
section 4. 

Finally 
the scalar potential can be written by combining all the components of
$A_1, A_2$ tensors  using the form of (\ref{V}) as
\bea
V & = & 
 - g^2 \left( \frac{3}{4} \left(2 |z_2|^2 +6 |z_1|^2 
\right) -\frac{2}{24}
 \left( 36 |y_1|^2 + 18 |y_2|^2+ 18 |y_3|^2 + 72 |y_4|^2+
24 |y_5|^2 \right) \right)  \nonu \\
& = & 2 g^2 c'^2 \left( \left(s^3+c^3 \right) s'^2 -3c \right) \nonu \\
& = & 
 \frac{1}{32} g^2 e^{-\frac{1}{\sqrt{2}} \la -2 \sqrt{2} \la'}
 \left(1+e^{\sqrt{2} \la'} \right)^2 
\left[ 3 + e^{2 \sqrt{2} \la} -30 e^{\sqrt{2} \la'} +3
e^{2\sqrt{2} \la'} \right. \nonu \\
& & \left. -24 e^{\sqrt{2} ( \la + \la')} -2 e^{
\sqrt{2} (2\la + \la')}+   e^{2\sqrt{2} (\la + \la')}\right]
\label{potential}
\eea
which is exactly the same form obtained by Warner \cite{warner} sometime ago
using $SU(8)$
coordinate system as an alternative approach. 

\begin{figure}[htb]
\label{fig2}
\vspace{0.5cm}
\epsfysize=8cm
\epsfxsize=8cm
\centerline{
\epsffile{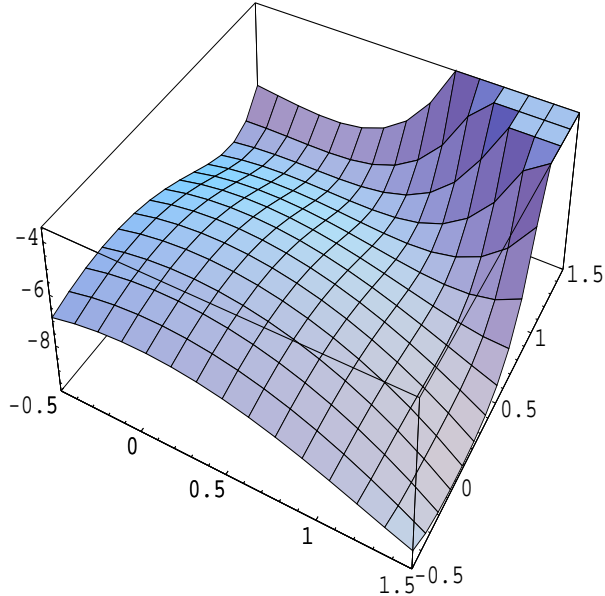} }
\vspace{0.5cm}
\caption{\sl Scalar potential $V(\la, \la')$. The left axis corresponds to
$\la$ and right one does $\la'$. The extremum value $V=-9\sqrt{3}/2=-7.79$
for $SU(3) \times U(1)$ occurs
around $\la=0.78$ and $\la'=0.93$ while the local maximum value 
$V=-6$ for $SO(8)$
appears around
$\la=0$ and $\la'=0$. We take $g^2$ as 1 for simplicity.  }
\end{figure}
\vspace{0.5cm}

The $AdS_4$-invariant ground-states correspond to $\la, \la'$ taking
constant values and the spacetime curvature maximally symmetric. The two 
vacua are as follows:
\bea
\begin{array}{|c|c|c|c|}
\hline
$\mbox{Gauge symmetry}$ 
& \la  & \la' & V \nonu \\
\hline
   SO(8) &  0  & 0 & - 6g^2 \nonu \\
\hline
 SU(3) \times U(1) &  \sqrt{2} \sinh^{-1} \left(\frac{1}{
\sqrt{3}} \right)=0.78  &  \sqrt{2} \sinh^{-1} \left(\frac{1}{
\sqrt{2}} \right)=0.93  & - \frac{9 \sqrt{3}}{2} g^2=-7.79 g^2 \nonu \\
\hline
\end{array}
\eea

Table 1. \sl Summary of two critical points: symmetry group,
vacuum expectation values of fields, and cosmological constants. \rm

The scalar potential $V(\la,\la')$ depicted in Figure 1 exhibits
the two critical points: $SO(8)$ point is a maximum point while $SU(3) 
\times U(1)$ is an other extremum point. 
The former is invariant under the full $SO(8)$
group while the latter is invariant only under the $SU(3) \times U(1)$ 
subgroup.

\section{ Three-Dimensional Super Conformal Field Theories}

In this section, by exploiting the results of section 2 on the Kaluza-Klein
spectrum under the deformation, we will find an operator that gives rise to
a renormalization group flow associated with  the symmetry breaking
$SO(8) \rightarrow SU(3) \times U(1) $ and get that the operator is relevant
at the $SO(8)$ fixed point but becomes irrelevant at the $SU(3) \times U(1)$
fixed point.  

\subsection{$SO(8)$ Invariant Conformal Fixed Point}

We will be identifying a renormalization group flow 
associated with symmetry breaking $SO(8) \rightarrow SU(3) \times U(1)$
in a three-dimensional strongly coupled field theory. We will show that 
the perturbation operator is relevant at the $SO(8)$ invariant UV fixed 
point corresponding to $OSp(8|4)$ extended supersymmetry 
but becomes irrelevant at the $SU(3) \times U(1)$ 
invariant IR fixed point corresponding to $OSp(2|4)$ extended supersymmetry.
To identify conformal field theory operator corresponding to the
perturbation while preserving $SU(3) \times U(1)$ symmetry, 
we will consider harmonic 
fluctuations of space-time metric and $\la(x)$ scalar field and 
$\la'(x)$ 
pseuodo-scalar field around 
$AdS_4 \times \S^7$. From the scalar potential Eq. (\ref{potential}),
one finds that the cosmological constant $\Lambda$ at $SO(8)$ fixed point
is given by
\bea
\La_{SO(8)}= - 6 g^2 \equiv -\frac{3}{r_{\rm UV}^2 \ell_{\rm pl}^2}  ,
\nonu
\eea
where $r_{\rm UV}$ is the radius of $AdS_4$ and $\ell_{\rm pl}$ 
is the eleven-dimensional Planck scale.
Conformal dimension of the perturbation operator representing this 
deformation
is calculated by fluctuation spectrum of the scalar and pseudo-scalar fields. 
From the scalar kinetic terms of 
$-|A_{\mu}^{\;\;\;IJKL}|^2/96$ and 
the explicit forms of $u^{IJ}_{\;\;KL}$ and $v^{IJKL}$ in 
Eq. (\ref{uv}),  
the resulting kinetic term turns out to be 
$-\frac{1}{2} \left[\frac{3}{4} 
(\partial_{\mu} \la)^2 +  (\partial_{\mu} \la')^2 
\right]$. 
After rescaling the $\lambda$ and $\la'$
fields as $\overline{\la}= \sqrt{\frac{3}{4}} \la, 
\overline{\la'}= \la'$, one finds that
the mass spectrum of the $\overline{\la}$ field around $SO(8)$ fixed point
in the unit of inverse radius of $AdS_4$ is given by:
\bea
M_{\overline{\la} \overline{\la}}^2(SO(8))  = 
 \left[ \frac{\pa^2 V}{\pa \overline{\la}^2}\right]_{
\overline{\la}=\overline{\la'}=0}  \,\,\, = \qquad  -4 g^2 \ell_{\rm pl}^2 
\,\,\, = \qquad -2 \frac{1}{r_{\rm UV}^2 }. 
\label{m^2}
\eea
Similarly,
the mass spectrum of the $\overline{\la'}$ field around $SO(8)$ fixed point
is given by:
\bea
M_{\overline{\la'} \overline{\la'}}^2(SO(8))  = 
 \left[ \frac{\pa^2 V}{\pa \overline{\la'}^2}\right]_{
\overline{\la}=\overline{\la'}=0}  \,\,\, = \qquad 
  -4 g^2 \ell_{\rm pl}^2 \,\,\,
= \qquad -2 \frac{1}{r_{\rm UV}^2 }. 
\label{M^2}
\eea

Via AdS/CFT correspondence, one finds that in the corresponding ${\cal N}=8$
superconformal field theory, the $SU(3) \times U(1)$ 
symmetric deformation ought to
be a relevant perturbation of conformal dimension $\De=1$ or $\De=2$.
Recall that, on $\S^7$, mass spectrum of the representation corresponding
to $SO(8)$ Dynkin label $\bf (n, 0, 2, 0)$ is given by 
$
\widetilde{M^2}= \left( (n+1)^2-9 \right) m^2
\nonu
$
where $m^2$ is mass-squared parameter of a given $AdS_4$ space-time and
a scalar field $S$ satisfies $(\De_{\mbox{AdS}}+\widetilde{M^2})S=0$.
This follows from the known mass formula \cite{bcers} $M^2=((n+1)^2-1) m^2$
for $0^{-(1)}$ and the fact that $M^2$ is traditionally defined according to
$(\De_{\mbox{AdS}}-8 m^2 +M^2)S=0$. For $\bf 35_c$ corresponding to $n=0$,
$\widetilde{M^2}_{\bf 35_{c}}=-8 m^2$ and this ought to equal to 
Eq.(\ref{M^2}).
Recalling that $r_{\rm UV}^2= r_{\S^7}^2/4=1/4m^2$,
\bea
 \left[ \frac{\pa^2 V}{\pa \overline{\la'}^2}\right]_{
\overline{\la}=\overline{\la'}=0}  \,\,\, =
 -2 \frac{1}{r_{\rm UV}^2 }= \widetilde{M^2}_{\bf 35_{c}}.
\nonu 
\eea
The conformal dimensions of the corresponding chiral operators
in the SCFT side are $\De = (n+4)/2$. Some of these operators may be 
identified with a product of two fermions times $n$ scalars.
Then  $\bf 35_c$ pseudo-scalars correspond to the conformal
primaries of $\De =2$ which consist of quadratic of Majorana gauginos in the
irreducible representations $\bf 8_c$ of $SO(8)$ of 3-dimensional
${\cal N}=8$ $SU(N_c)$ gauge theory living on the worldvolume of $N_c$
coincident M2 branes. 

On the other hand, the mass spectrum of the representation corresponding
to $SO(8)$ Dynkin label $\bf (n+2, 0, 0, 0)$ is given by $
\widetilde{M^2}= \left( (n-1)^2-9 \right) m^2$.
This follows from the known mass formula \cite{bcers} $M^2=((n-1)^2-1) m^2$
for $0^{+(1)}$.
 For $\bf 35_v$ corresponding to $n=0$,
$\widetilde{M^2}_{\bf 35_{v}}=-8 m^2$ and this ought to equal to
Eq. (\ref{m^2}):
\bea
 \left[ \frac{\pa^2 V}{\pa \overline{\la}^2}\right]_{
\overline{\la}=\overline{\la'}=0}  \,\,\, =
 -2 \frac{1}{r_{\rm UV}^2 }= \widetilde{M^2}_{\bf 35_{v}}.
\nonu 
\eea
The conformal dimensions of the corresponding chiral operators
in the SCFT side are $\De = (n+2)/2$. Some of these operators may be 
identified with a product of $(n+2)$'s scalar fields in the vector
multiplet.
The $\bf 35_v$ scalars correspond to the conformal primaries 
of $\De =1$ which consist of quadratic of real scalars in the 
irreducible representation $\bf 8_v$ of $SO(8)$ of 3-dimensional
${\cal N}=8$ $SU(N_c)$ gauge theory.

\subsection{$SU(3) \times U(1)$ Invariant Conformal Fixed Point}

Let us next consider the conformal fixed point corresponding to the 
$SU(3) \times U(1)$ symmetry. 
Again, from the scalar potential Eq.(\ref{potential}),
one finds that cosmological constant $\Lambda$ at $SU(3) \times U(1)$ fixed
point is given by
\bea
\La_{SU(3) \times U(1)}= -  \frac{9 \sqrt{3}}{2}
g^2  \equiv -\frac{3}{r_{\rm IR}^2 \ell_{\rm pl}^2}.  
\nonu
\eea
One calculates mass spectrum of the scalar and pseuodo-scalar fields 
straightforwardly:
\bea
M_{\overline{\la} \overline{\la}}^2(SU(3) \times U(1)) & = 
& \left[ \frac{\pa^2 V}{\pa \overline{\la}^2}\right]_{
\overline{\la_{\rm ext}},
\overline{\la'_{\rm ext}}} \,\,\, = \qquad  
 3 \sqrt{3} g^2 \ell_{\rm pl}^2 , \nonu \\
M_{\overline{\la} \overline{\la'}}^2(SU(3) \times U(1)) & = 
& \left[ \frac{\pa^2 V}{\pa \overline{\la} 
\pa \overline{\la'}} \right]_{
\overline{\la_{\rm ext}}, \overline{\la'_{\rm ext}}} 
= \qquad 6 \sqrt{3} g^2 \ell_{\rm pl}^2,    \nonu \\
  M_{\overline{\la'}\overline{\la'}}^2(SU(3) \times U(1)) & = & 
\left[ \frac{\pa^2 V}{\pa \overline{\la'}^2}\right]_{
\overline{\la_{\rm ext}},
\overline{\la'_{\rm ext}}} \,\,\, =  \qquad 6 \sqrt{3} 
g^2 \ell_{\rm pl}^2,  \nonu 
\eea
where $\overline{\la_{\rm ext}}$ and $\overline{\la'_{\rm ext}}$ takes the form
of vacuum expectation values in Table 1.
\bea
\sinh \left(\sqrt{\frac{2}{3}} \overline{\la_{\rm ext}} \right) = 
\frac{1}{\sqrt{3}}, \qquad
\sinh \left(\frac{1}{\sqrt{2}} \overline{\la'_{\rm ext}} \right) = 
\frac{1}{\sqrt{2}}. \nonu
\eea
Diagonalizing the mass matrix , one obtains the mass eigenvalues as follows: 
\bea
M^2 = \frac{3}{2} \left( 3- \sqrt{17} \right) \times \sqrt{3}
g^2 \ell_{\rm pl}^2  , \quad
  \frac{3}{2} \left( 3+ \sqrt{17} \right) \times \sqrt{3}
g^2 \ell_{\rm pl}^2.
\nonu
\eea

One finds that
the fluctuation spectrum for $\la, \la'$ fields around  
$SU(3) \times U(1)$ fixed point
takes one negative value and one positive value:
\bea
M_{\widetilde{\la} \widetilde{\la}}^2(SU(3) \times U(1)) =    
-\left( \sqrt{17} -3 \right) \frac{1}{r_{\rm IR}^2}, \qquad 
  M_{\widetilde{\la'}\widetilde{\la'}}^2(SU(3) \times U(1))  =  
 \left( 3+\sqrt{17} \right) \frac{1}{r_{\rm IR}^2},
\nonu
\eea
where $\widetilde{\la} =\frac{1}{\sqrt{13}}(2\overline{\la}+
3\overline{\la'})$ 
and $\widetilde{\la'}= \frac{1}{\sqrt{13}}(3\overline{\la}-2\overline{\la'})$.
Under $SO(8) \rightarrow SU(3) \times U(1)$, the branching rule
of a $SO(8)$ Dynkin label $\bf (0, 0, 2, 0) \oplus (2, 0, 0, 0)$ corresponding
to the representation $\bf 35_c \oplus 35_v$ in terms of
$SU(3)$ representation is given as follows:
\bea
\bf 70 & = & \bf 1(0) \oplus 1(0) \oplus 1(1) \oplus 1(0) 
\oplus 1(-1) \oplus 8(0) \oplus 8(0) \nonu \\
& & \bf \oplus 3(-1/3) \oplus \bar{3}(1/3) \oplus 6(1/3) \oplus 6(-2/3) \oplus
\bar{6}(-1/3) \oplus \bar{6}(2/3) \nonu \\
& & \bf 
\oplus 3(2/3) \oplus 3(-1/3) \oplus 3(-1/3) \oplus \bar{3} (-2/3) \oplus
\bar{3}(1/3) \oplus \bar{3}(1/3) \oplus 1(0) \nonu
\eea
where the number in the brackets after $SU(3)$ representation is the
hypercharge, Y, of it.
Since the deformation preserves $SU(3) \times U(1)$ group, the spectrum
ought to correspond to that of the singlet.

From the above mass spectrum, one finds that, in ${\cal N}=2$ superconformal 
field theory, the $SU(3) \times U(1)$ symmetric 
deformation ought to be an irrelevant perturbation 
of conformal dimension $\De=(3+\sqrt{21+4\sqrt{17}})/2=4.5616$.
The corresponding eigenvector determines the direction from which the flow
approaches the fixed point. The irrelevant operator 
in the field theory that controls this flow has this dimension.
We thus conclude that the perturbation operator dual to the $\widetilde{
\lambda'}$
field induces nontrivial renormalization group flow from ${\cal N}=8$
superconformal UV fixed point with $SO(8)$ symmetry to ${\cal N}=2$ 
superconformal IR fixed point with $SU(3) \times U(1)$ symmetry.
 Scaling dimension of other 
deformations is $\Delta = (3 \pm \sqrt{21-4\sqrt{17}})/2$ 
for $\widetilde{\la}$ field.

\section{ Supersymmetric Domain Wall and RG Flows}

In this section, we investigate domain walls arising in supergravity
theories with a nontrivial superpotential defined on a restricted 
2-dimensional slice of the scalar manifold.
On the subsector, one can write the supergravity potential in the
canonical form.
One of the eigenvalues of $A_1^{\;\;IJ}$ tensor (\ref{a1}), $z_2$
provides a ``superpotential'' $W$ related to
scalar potential $V$ by
\bea
V(\la, \la')= g^2 \left[ \frac{16}{3}  \left(\frac{\partial W}{\partial 
\la} \right)^2 + 4   \left(\frac{\partial W}{\partial 
\la'} \right)^2 - 6  W^2 \right]
\label{pot}
\eea
where
\bea
W(\la, \la')  =  \frac{1}{16} e^{-\frac{1}{2\sqrt{2}} \la -\sqrt{2} \la'}
 \left( 3 - e^{\sqrt{2} \la} + 6 e^{\sqrt{2} \la'} +3
e^{2\sqrt{2} \la'}+6 e^{\sqrt{2} ( \la + \la')} -e^{
\sqrt{2} (\la +2 \la')} \right).
\eea
Note that superpotential $W$ is real rather than complex and this 
fact will make finding a BPS solution easier. 
At the two critical points, 
the gradients of $W$ with respect to $\la, \la'$ vanish. That is,
supersymmetry preserving extrema of the potential satisfy
$\frac{\partial W}{\partial \la}| = \frac{\partial W}{ \partial \la'}| =0$.
This implies that supersymmetry preserving vacua have negative cosmological
constant:the scalar potential $V$  at the two critical points becomes
$V=-6 g^2 W^2$ in very simple form.
The superpotential $W$ has the following values at the 
two critical points yielding stable $AdS_4$ vacua.
\bea
\begin{array}{|c|c|c|c|}
\hline
$\mbox{Gauge symmetry}$ 
& \la  & \la' & W \nonu \\
\hline
   SO(8) &  0  & 0 & 1 \nonu \\
\hline
 SU(3) \times U(1) &  \sqrt{2} \sinh^{-1} \left(\frac{1}{
\sqrt{3}} \right)=0.78  &  \sqrt{2} \sinh^{-1} \left(\frac{1}{
\sqrt{2}} \right)=0.93  & \frac{3^{3/4}}{2} \nonu \\
\hline
\end{array}
\eea

Table 2. \sl Summary of two critical points in the context of
superpotential : symmetry group,
vacuum expectation values of fields, and superpotential. The 
superpotential $W(\la, \la')$ also exhibits the two critical points:
$SO(8)$ point is a minimum while $SU(3) \times U(1)$ point is 
an other extremum. \rm

To construct the superkink corresponding  to 
the supergravity description of the nonconformal
RG flow from one scale to another 
 connecting the above two critical points,
the form of a 3d Poincare invariant metric but breaking 
the full conformal group $SO(3,2)$ invariance takes the form:
\bea
ds^2= e^{2A(r)} \eta_{\mu \nu} dx^{\mu} dx^{\nu} + dr^2, \;\;\;
\eta_{\mu \nu}=(-,+,+)
\label{ansatz}
\eea
characteristic of space-time with a domain wall where $r$ is the
coordinate transverse to the wall.
By change of variable $U(r)=e^{A(r)}$ at the critical points,
the geometry becomes $AdS_4$ space with a cosmological constant
$\La$ equal to the value of $V$ at the critical points: 
$\La= -3 (\pa_r A)^2$. In the dual theory this corresponds to  
a superconformal fixed point of the RG flow.
We are looking for solutions that are asymptotic
to $AdS_4$ space both for $\la, \la' \rightarrow 0$ for $r \rightarrow 
\infty $ so that the background is asymptotic to the ${\cal N}=8$  
supersymmetric $AdS_4$ background at infinity with 
$r_{UV}$ 
while $\la \rightarrow \la_{IR}= \sqrt{2} \sinh^{-1} \left(\frac{1}{
\sqrt{3}} \right), \la' \rightarrow \la'_{IR}= \sqrt{2} \sinh^{-1} 
\left(\frac{1}{
\sqrt{2}} \right)$ for
$r \rightarrow -\infty$ with $r_{IR}$ and 
so we approach a new conformal fixed point.
The second order differential 
equations of motion for the scalars and the metric from (\ref{action}) read
\bea
& & \frac{\partial^2 \la}{\partial r^2} +3 \left( \frac{d A}{d r}
\right) \left( 
\frac{\partial \la}{\partial r} \right) 
 =  \frac{4}{3} \frac{\partial V}{
\partial \la}, \nonu \\
& & \frac{\partial^2 \la'}{\partial r^2} +3 \left( \frac{d A}{d r}
\right) \left( 
\frac{\partial \la'}{\partial r} \right)  
 =   \frac{\partial V}{
\partial \la'}, \nonu \\
& & 6 \left( \frac{d A}{d r} \right)^2-
 \frac{3}{4} 
\left( \frac{\partial 
\la }{\partial r} \right)^2
- \left( \frac{\partial 
\la'}{\partial r} \right)^2  
+2 V  =  0, \nonu \\
& & 4  \frac{d^2 A}{d r^2} +
6 \left( \frac{d A}{d r} \right)^2+
 \frac{3}{4} 
\left( \frac{\partial 
\la}{\partial r} \right)^2
+
\left( \frac{\partial 
\la'}{\partial r} \right)^2 
+2 V  =  0.
\label{eom}
\eea
The last relation can be obtained by differentiating the third one and using
other relations. 
Only three of them are independent.
By substituting the domain wall ansatz (\ref{ansatz})
into the Lagrangian (\ref{action}), 
the Euler-Lagrangian equations are the first, second and fourth equations of
(\ref{eom}) for the functional $E[A, \la, \la']$ 
\cite{st} with the integration by parts
on the term of $\frac{d^2 A}{d r^2}$ where
\bea
E[A,\la,\la']  
& = &  -\frac{1}{2} \int_{-\infty}^{\infty} dr e^{3A} \left[
 -6 \left( \frac{d A}{d r} \right)^2 
+\frac{3}{4} \left( \frac{\partial 
\la}{\partial r} \right)^2 + \left( \frac{\partial \la'}
{\partial r} \right)^2   +2V  \right] \nonu \\
&  = &  -\frac{1}{2} \int_{-\infty}^{\infty} dr e^{3A} \left[
 -6 \left( \frac{d A}{d r} +  \sqrt{2} g W \right)^2 
+\frac{3}{4} \left( \frac{\partial 
\la}{\partial r}- \frac{8}{3} \sqrt{2} g  \frac{\partial W}{\partial 
\la}\right)^2 \right. \nonu \\
& & \left. + \left( \frac{\partial \la'}
{\partial r}- 2 \sqrt{2} g  \frac{\partial W}{\partial \la'} 
\right)^2 \right] - 
2 \sqrt{2} g e^{3A} W |_{-\infty}^{\infty} \nonu \\
& &   \geq - 
2 \sqrt{2} g \left(  e^{3A} W(\infty) -e^{3A} W(-\infty) \right)   
\label{Ebound}
\eea
which is so-called topological charge or domain wall number.
As $r$ goes from large positive values(the UV) to large negative 
values(the IR) the change in $e^{3A} W$ is a measure of the topological
charge of the superkink.

Then $E\left[A, \la, \la'\right]$ 
is extremized by the following so-called
BPS domain wall solutions \footnote{In \cite{cveticetal}, static domain 
wall solutions in ungauged ${\cal N}=1$ supergravity theories were found. }:
\bea
\frac{\partial \la}{\partial r}  = \pm  
\frac{8}{3} \sqrt{2} g  \frac{\partial W}{\partial \la}, \qquad
\frac{\partial \la'}{\partial r}  = \pm  
2 \sqrt{2} g  \frac{\partial W}{\partial \la'}, \qquad
\frac{d A}{d r}  =  \mp \sqrt{2} g W.
\label{first}
\eea
It is evident that the left hand sides of the first and second relations 
vanish as one approaches the supersymmetric extrema, i.e. 
$\frac{\partial W}{\partial \la}| = \frac{\partial W}{ \partial \la'}| =0$
thus indicating a domain wall configuration.
The asymptotic behaviors of $A(r)$ are $A(r) \rightarrow r/r_{UV} + const$
for $r \rightarrow \infty$ and  
$A(r) \rightarrow r/r_{IR} + const$
for $r \rightarrow -\infty$. Then by differentiating $A(r)$ with respect to
$r$ those of $\pa_r A$ become 
$\pa_r A \rightarrow 1/r_{UV}$
for $r \rightarrow \infty$ and  
$\pa_r A \rightarrow 1/r_{IR}$.
At the two critical points, since $V =-6 g^2 W^2$,  one can write 
the inverse radii of $AdS_4$ as cosmological costant or superpotential $W$.
Therefore we conclude that $1/r$ is equal to $\pm \sqrt{2} g W$. This fact
is encoded in the last equation of (\ref{first}).
It is straightforward to verify that (\ref{first}) 
satisfy the gravitational and
scalar equations of motion given by
second order differential equations (\ref{eom}).
Using (\ref{first}), the monotonicity \cite{domainwall} of 
$\frac{d A}{d r}$  which is related to the local potential energy
of the superkink leads to
\bea
\frac{d^2 A}{d r^2} = -4 g^2 \left( \frac{4}{3} \left( \frac{\partial W
}{\partial \la} \right)^2 +\left( \frac{\partial W}{\partial \la'} 
\right)^2 \right) = - \left( \frac{3}{8} \left( 
\frac{\partial \la}{ \partial r} \right)^2+
\frac{1}{2} \left( \frac{\partial \la'}{\partial r} \right)^2 \right) \leq 0.
\label{mono}
\eea
One can understand the above bound (\ref{Ebound}) 
as a conseqence of supersymmetry preserving
bosonic background. 
In order to find supersymmetric bosonic backgrounds, the variations
of spin $1/2, 3/2$- fields should vanish. 
From \cite{dn2}, the gravitational and scalar parts of these variations are:
\bea
\de\psi_{\mu}^i & =& 
 2 D_{\mu} \epsilon^i - \sqrt{2} g A_1^{\;ij} \ga_{\mu} \ep_j, 
\nonu \\
\de \chi^{ijk} & = & - \ga^{\mu} A_{\mu}^{\;\;ijkl} \ep_l- 2 g 
A_{2l}^{\;\;\;ijk} 
\ep^l, \nonu
\eea
where 
\bea
 D_{\mu} \epsilon^i = \partial_{\mu} \ep^i - \frac{1}{2} 
\omega_{\mu a b} \sigma^{ab} \ep^i
+ \frac{1}{2} 
{\cal B}_{\mu\;\;\;j}^{\;\;i} \ep^j, \;\;\; 
{\cal B}_{\mu\;\;\;j}^{\;\;i} = \frac{2}{3} \left( u^{ik}_{\;\;\;IJ}
\partial_{\mu} u_{jk}^{\;\;\;IJ} - v^{ikIJ} \partial_{\mu} v_{jkIJ} \right)
\nonu
\eea
where $\omega$ is a spin connection, $\sigma$ a commutator of two gamma 
matrices and ${\cal B}$ is a $SU(8)$ gauge field for local 
$SU(8)$ invariance of the theory.
In order to make the AdS/CFT correspondence completely, one should 
find the flow between $SO(8)$ fixed point and the $SU(3) \times U(1)$ 
fixed
point. One should be able to preserve ${\cal N}=2$ supersymmetry on the
branes all along the flow. The vanishing of $ \de \chi^{ijk}$ associates the 
derivatives of scalar $\la$ and  pseudo-scalar $\la'$ 
with respect to $r$ with the gradients of
superpotential $W$. 
The variation of 56 Majorana spinors $\chi^{ijk}$ gives rise to
the first order differential equations of $\la$ and $\la'$ by exploiting
the explicit forms of $ A_{\mu}^{\;\;ijkl}$ and $A_{2l}^{\;\;\;ijk}$ in the
appendix.
Although there is a summation over the last index $l$ appearing in
$ A_{\mu}^{\;\;ijkl}$ and $A_{2l}^{\;\;\;ijk}$, nonzero contribution
runs over only one index. When $i=1, j=2, k=7$ 
and $l=8$, the variation of  $\chi^{127}$ leads to 
\bea
\de \chi^{127}=
\frac{1}{2} \frac{\partial \la}{\partial r}  \ep_8 -
 2g y_2 \ep^8 =\frac{1}{2} \frac{\partial \la}{\partial r}  \ep_8
- \frac{4\sqrt{2}}{3} g \frac{\partial W  }{\partial \la } \ep^8
\nonu
\eea  
where we used the fact that $y_2$ in (\ref{yis}) 
can be written as gradient of 
superpotential.
From this, we arrive at the first equation of BPS domain wall solutions
(\ref{first}).
Similarly, when $i=1, j=6, k=3$ 
and $l=8$, the variation of  $\chi^{163}$ leads to 
\bea
\de \chi^{163}=
-\frac{i}{2} \frac{\partial \la'}{\partial r}  \ep_8 -
 2g y_5 \ep^8 =-\frac{i}{2} \frac{\partial \la'}{\partial r}  \ep_8
+ \sqrt{2} i g \frac{\partial W  }{\partial \la' } \ep^8. 
\nonu
\eea  
The vanishing of this is exactly same as the second equation of (\ref{first}).
One can check also there exist similar BPS domain wall solutions for nonzero
$\ep_7$.
From these first order differential equations, it is straightforward to check
with the help of appendix that all other 
supersymmetric parameters $\ep_i$ where $i=1, \cdots, 6$ vanish. 
As a result, the flow preserves ${\cal N}=2$ supersymmetry, generated by
$\ep_7$ and $\ep_8$, on the M2 brane. 
Moreover,
the variation of gravitinos $\psi^{i}_{r}$ with vanishing 
${\cal B}_{\mu\;\;\;j}^{\;\;i}$ 
will lead to
\bea
\de\psi_{r}^8 = -\frac{d A}{d r} \ep^8 - \sqrt{2} g W \ep_8.
\nonu
\eea
which will also produce the third equation of (\ref{first}). 
Now we have shown that there exists a supersymmetric flow if and only if 
the equations (\ref{first}) are satisfied, that is, the flow is determined
by the steepest descent of the superpotential and the cosmology $A(r)$ is
determined directly from this steepest descent.

Let us consider mass, $\widetilde{M^2}$ for the $\overline{\la},
\overline{\la'}$ at the
critical points of superpotential $W$. 
By differentiating (\ref{pot}) and putting
$\frac{\partial W}{\partial \overline{\la}}|=\frac{\partial W}{\partial 
\overline{\la'}}|= 0$, we get
\bea
\widetilde{M}^2  & = & 2 g^2 W^2
\left(
\begin{array}{cc}
{\cal U_{\overline{\la} \overline{\la}}} 
\left( {\cal U_{\overline{\la} \overline{\la}}} -3 \right) +
{\cal U_{\overline{\la} \overline{\la'}}} 
 {\cal U_{\overline{\la'} \overline{\la}}} & 
  {\cal U_{\overline{\la} \overline{\la'}}} 
\left( {\cal U_{\overline{\la'} \overline{\la'}}} -3 \right) +
{\cal U_{\overline{\la} \overline{\la}}} 
 {\cal U_{\overline{\la} \overline{\la'}}}  \nonu \\
 {\cal U_{\overline{\la} \overline{\la'}}} 
\left( {\cal U_{\overline{\la'} \overline{\la'}}} -3 \right) +
{\cal U_{\overline{\la} \overline{\la}}} 
 {\cal U_{\overline{\la} \overline{\la'}}} 
&  {\cal U_{\overline{\la'} \overline{\la'}}} 
\left( {\cal U_{\overline{\la'} \overline{\la'}}} -3 \right) +
{\cal U_{\overline{\la'} \overline{\la}}} 
 {\cal U_{\overline{\la} \overline{\la'}}} 
\end{array}
\right), \nonu \\ 
& = & \frac{3 \sqrt{3}}{2} g^2 
\left(
\begin{array}{cc}
 2 & 4 \nonu \\
4 & 4 
\end{array} \right) \nonu
\eea
where
\bea
{\cal U_{\overline{\la} \overline{\la}}}= \frac{2}{W} 
\left( \frac{\partial^2 W}{ \partial \overline{\la}^2} \right), \qquad
{\cal U_{\overline{\la} \overline{\la'}}}= \frac{2}{W} \left( 
\frac{\partial^2 W}{ \partial \overline{\la} \partial \overline{\la'}} 
\right), \qquad
{\cal U_{\overline{\la'} \overline{\la'}}}
=\frac{2}{W} \left(
\frac{\partial^2 W}{ \partial \overline{\la'}^2} \right).
\nonu
\eea
The mass scale is set by the inverse radius, $1/r$, of the $AdS_4$ space
and this can be written as $1/r=\ell_p \sqrt{-V/3}=\sqrt{2} g 
 W  $ where we used $V=-6g^2  W^2$. Via AdS/CFT
correspondence, ${\cal U}$ is related to the conformal dimension $\De$ of
the field theory operator dual to the fluctuation of the fields 
$\overline{\la}, \overline{\la'}$. Since the matrix ${\cal U}$ is real
and symmetric, it has real eigenvalues $\de_k$ and the eigenvalues of
$\widetilde{M}^2 r^2 $ are given by $\de_k \left( \de_k -3 \right)$. 
Since a new radial coordinate $U(r)=e^{A(r)}$ is 
the renormalization group scale on the flow, we should
find the leading
contributions to the $\be$ functions of the couplings $\overline{\la},
\overline{\la'}$ 
in the neighborhood of the end points of the flow. 
At a fixed point, the fields are constants and corresponding $\beta$ functions
vanish.
Since $\frac{d}{d r}=
\frac{d A}{d r} U \frac{d}{d U}=-\sqrt{2} g W U \frac{d}{d U}$, (\ref{first})
becomes 
\bea
U \frac{d}{d U} \overline{\la} & = & - \frac{2}{W}  \frac{\partial W}{  
\partial \overline{\la}}  \approx - 
\left( {\cal U}_{\overline{\la} \overline{\la}}|
\de \overline{\la} + {\cal U}_{\overline{\la} \overline{\la'}}|
\de \overline{\la'} \right), \nonu \\
U \frac{d}{d U} \overline{\la'} & = & - \frac{2}{W}  \frac{\partial W}{  
\partial \overline{\la'}}  \approx - 
\left( {\cal U}_{\overline{\la'} \overline{\la}}| 
\de \overline{\la} + {\cal U}_{\overline{\la'} \overline{\la'}}|
\de \overline{\la'} \right), \nonu 
\eea
where we expanded to first order in the neighborhood of a critical point.
Thus ${\cal U}$ determines the behavior of the scalar $\overline{\la},
\overline{\la'}$
near the critical points. 
The RG flow of the coupling costants of the field theory is encoded 
in the $U$ dependence of the fields. 
To depart the UV fixed point($U=+ \infty$) the 
flow must take place in directions in which the operators must be relevant
and to approach the IR fixed point($U \rightarrow 0$) the corresponding 
operators must be irrelevant.  

The contour maps of $V$ and $W$ on the $(\la, \la')$ parameter space are
shown in Figure 2. The map of $V$ shows two extrema. At $(\la, \la')=(0, 0)$,
it is the maximally supersymmetric  and locally maximum of $V$ while minimum
of $W$. At $(\la, \la')=(0.78, 0.93)$, it is ${\cal N}=2$ supersymmetric and
other extremum of both $V$ and $W$. 
A numerical solution of the steepest descent equation connecting these two 
critical points can be
obtained numerically. 

By realizing the fact that the scalar potential (\ref{pot}) has a symmetry
of $W \rightarrow -W$, the BPS domain wall solutions with $-W$ also 
satisfy the minimization condition of energy (\ref{Ebound}) 
and satisfy equations of motion. 
By taking the opposite signs in the right hand sides of (\ref{first}) and 
differentiating $W$ with respect to $\la$ and $\la'$, one gets $\la(r),
\la'(r)$ and $\pa_r A(r)$ which interpolate between the two
supersymmetric vacua in Figure 3 numerically. 
By change of variable,
$\tanh r$ rather than $r$, one draws them between $-1(r=-\infty)$ and 
$1(r=\infty)$ in the horizontal
axis. Starting IR fixed point, with initial data 
$\la =0.78, \la'=0.93( \mbox{or}\;\; 
e^{\frac{\la}{2\sqrt{2}}}=1.32, e^{\frac{\la'}{2\sqrt{2}}}=1.39)$, they 
decrease monotonically and finally go to the expectation values, 
$\la =0 =\la'( \mbox{or} \;\; e^{\frac{\la}{2\sqrt{2}}} =1= 
e^{\frac{\la'}{2\sqrt{2}}})$ of UV fixed point. 
Similarly,    
starting IR fixed point, with initial condition
$\pa_r A = \sqrt{2} g W = 2.74$ for $g=1.7$, it 
decreases monotonically showing the property of
(\ref{mono}) and finally goes to the expectation values, 
$\pa_r A = 2.40 $ of UV fixed point.

In summary, we have found an operator that gives rise to RG flow 
related to the symmetry breaking $SO(8) \rightarrow SU(3) \times U(1)$
 and got that the operator is relevant at the $SO(8)$ fixed point but 
irrelevant at the $SU(3) \times U(1)$ fixed point. The ability of
writing de Wit-Nicolai scalar potential in terms of superpotential
allowed us to determine BPS domain wall solutions easily. This superpotential
originates from the structure of contracted T-tensor which was a cubic
in 28-beins $u$ and $v$. From known supersymmetry variation of spin 1/2, 3/2
fields in gauged ${\cal N}=8$ supergravity, we were able to verify
that supersymmetry preserving bosonic background in supergravity theory
results in BPS domain wall solutions. The leading contributions to the 
$\be$ functions of the couplings of the field theory were encoded in
a quantity, second derivatives of superpotential with respect to scalar
and pseudo-scalars divided by superpotential itself. 

So far, various critical
points of gauged ${\cal N}=8$ supergravity potential are known for at least 
$SU(3)$ invariance \cite{warner}. It would be interesting to study the 
possiblity of existence of other critical points  of 
gauged ${\cal N}=8$ supergravity by requiring smaller symmetry group 
invariance.
  
\begin{figure}[htb]
\label{fig1}
\vspace{0.5cm}
\epsfysize=6cm
\epsfxsize=6cm
\centerline{
\epsffile{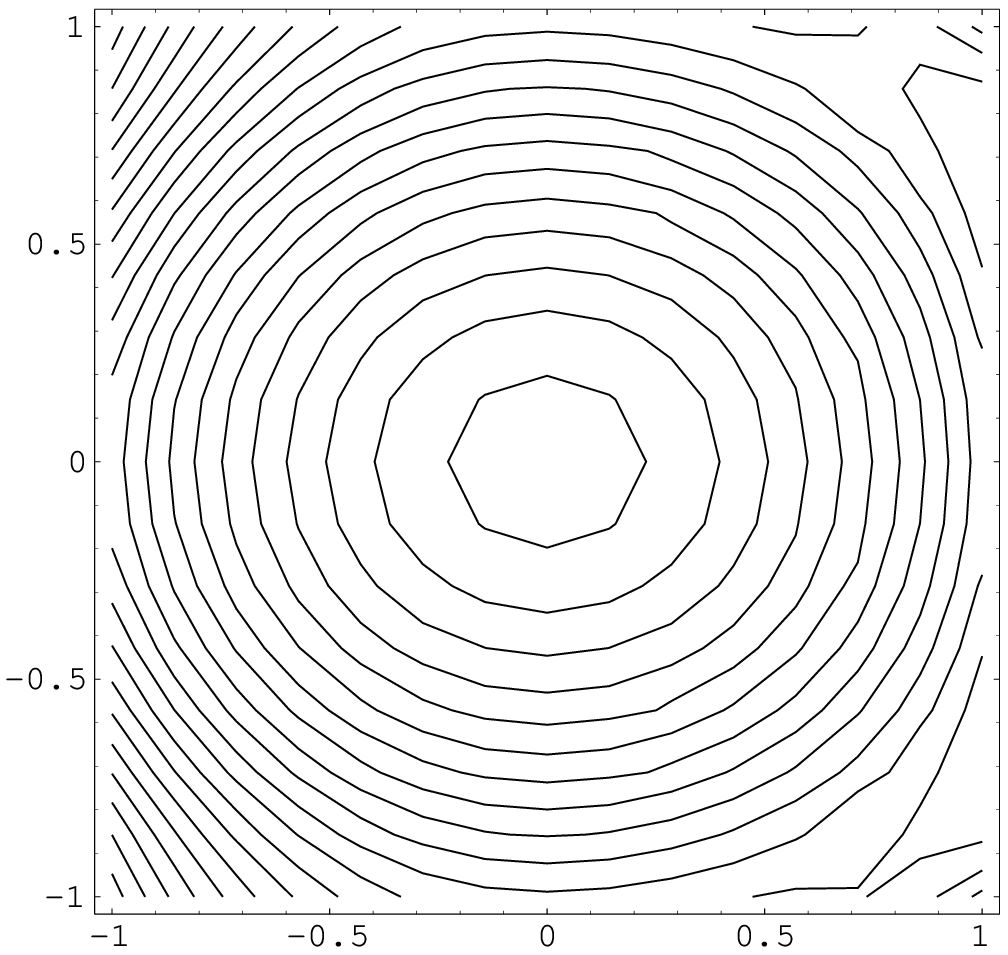}  
\epsfysize=6cm
\epsfxsize=6cm
\epsffile{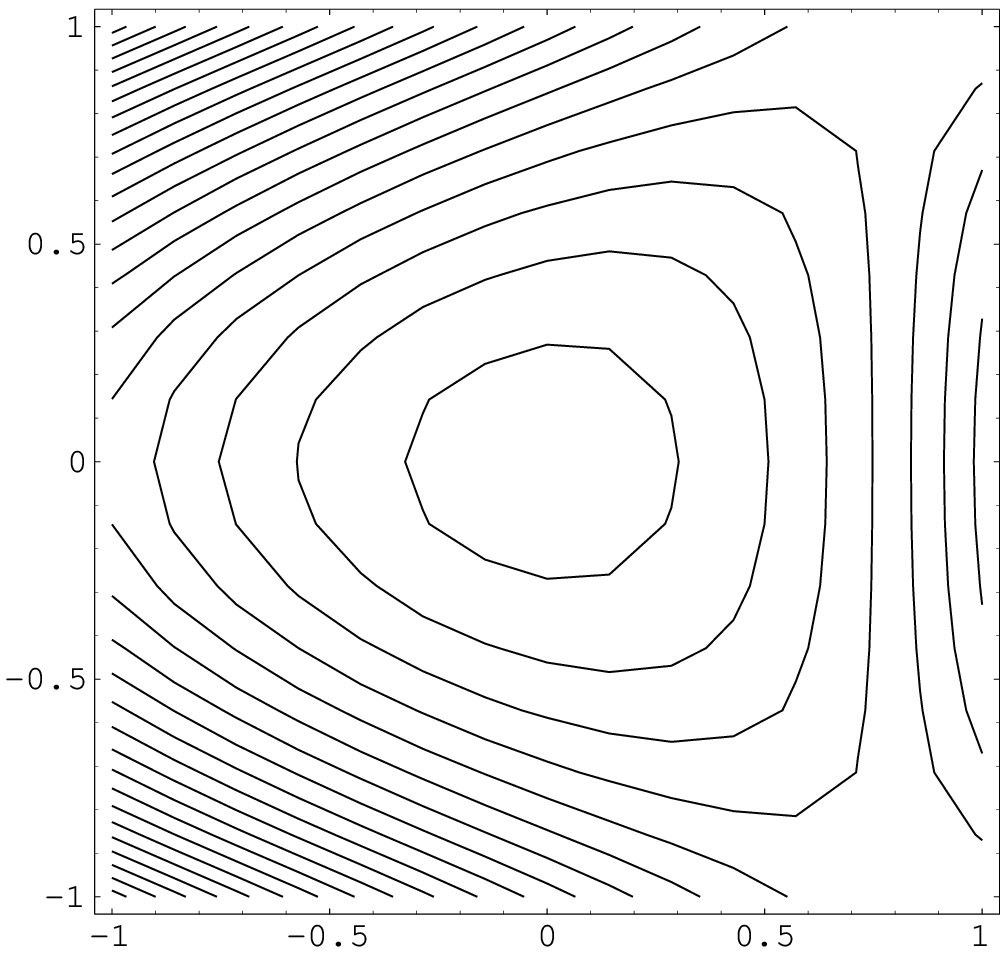}  }
\vspace{0.5cm}
\caption{\sl 
The contour map of $V$ (on the left) and $W$ (on the right), with
$\la'$ on the vertical axis and $\la$ on the horizontal axis.  $V$ has
vanishing first derivatives in all directions orthogonal to the plane.
At $(\la, \la')=(0, 0)$,
it is the maximally supersymmetric  and locally maximum of $V$ while minimum
of $W$. At $(\la, \la')=(0.78, 0.93)$, it is ${\cal N}=2$ supersymmetric and
other extremum of both $V$ and $W$.}
\end{figure}
\vspace{0.5cm}

\begin{figure}[htb]
\label{fig3}
\vspace{0.5cm}
\epsfysize=5cm
\epsfxsize=5cm
\centerline{
\epsffile{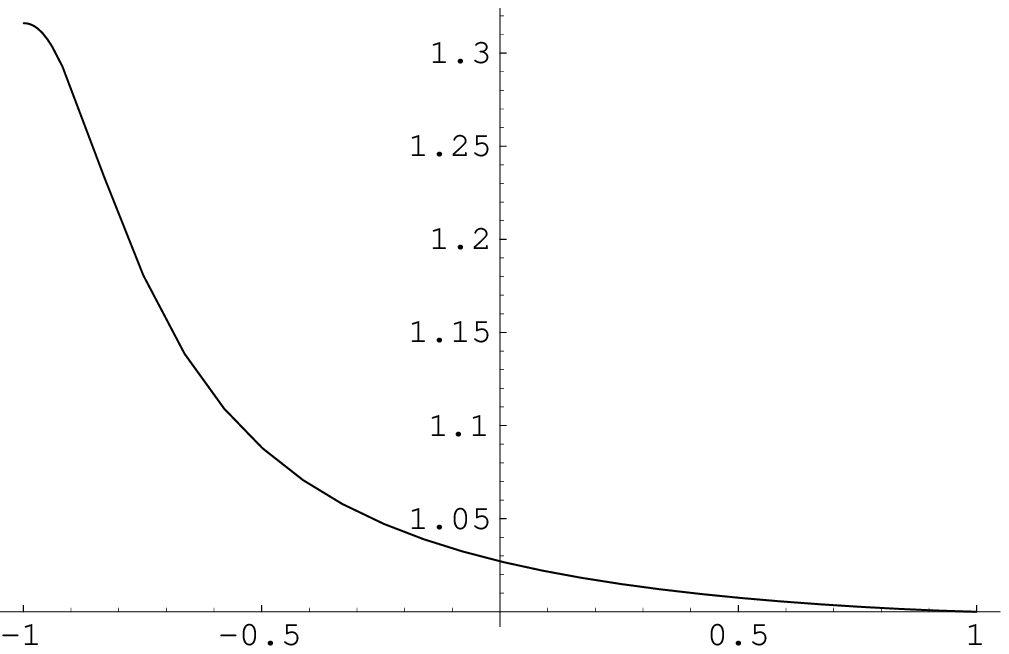}  
\epsfysize=5cm
\epsfxsize=5cm
\epsffile{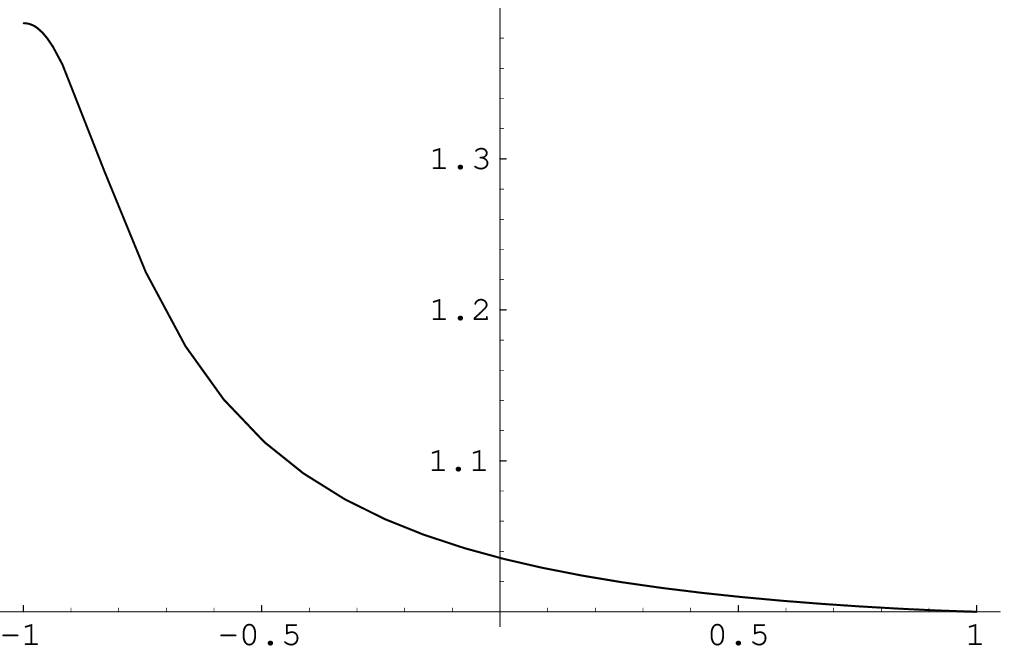}  
\epsfysize=5cm
\epsfxsize=5cm
\epsffile{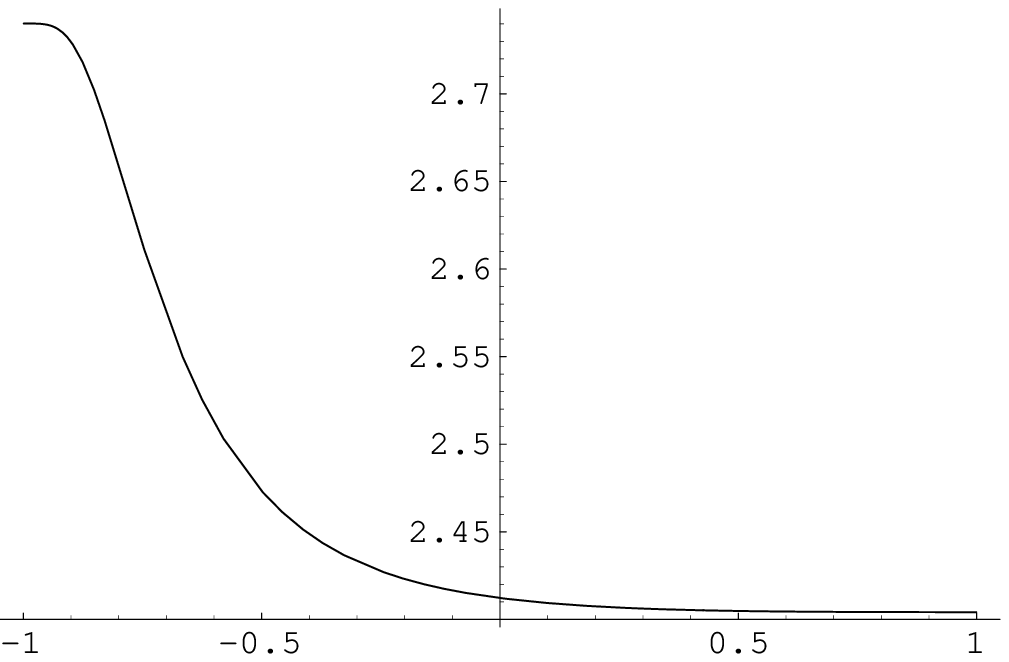}  }
\vspace{0.5cm}
\caption{\sl 
The plots of $e^{\frac{\la}{2\sqrt{2}}}$ (on the left), 
$e^{\frac{\la'}{2\sqrt{2}}}$ (on the middle) and  $\pa_r A$ (on the right), 
with
$\tanh r$ on the horizontal axis. They arrive at 1, 1, 2.40 respectively
 when $r \rightarrow \infty$ or $\tanh r|_{r \rightarrow \infty} =1$.
The value of $\la =0=\la'$ 
corresponds to the expectation values of fields of UV fixed point.  
The asymptotic value of last one  
is consistent with the vlaue of $\pa_r A = \sqrt{2} g W $
at UV fixed point where $W  =1$. 
We took $g=1.7$ for simplicity.     }
\end{figure}
\vspace{0.5cm}

\section{Appendix }
The 28-beins $u$ and $v$ fields can be obtained 
by exponentiating the vacuum expectation values $\phi_{ijkl}$:
\bea
u^{IJ}_{\;\;\;KL} & = & \mbox{diag}
\left(
 u_ 1, u_2, u_3, u_4, u_5, u_6, u_7 \right), \nonu \\
v^{IJKL} & = & \mbox{diag}
\left(
 v_ 1, v_2, 
v_3, v_4, v_5, v_6, v_7
\right),
\nonu
\eea
where each submatrix is $4 \times 4$ matrix and we denote the
antisymmetric index pairs $[IJ]$ and $[KL]$ explicitly for convenience.
\bea
&& u_1 = \left(
\begin{array}{ccccc}
& \left[12 \right] & \left[34 \right] & \left[56 \right] & 
\left[78 \right] \nonu \\
 \left[12 \right] &  A & B & B &  B \nonu \\
\left[34 \right]  & B & A & B & 
B \nonu \\
 \left[ 56\right] & B & B & A & 
B \nonu \\
 \left[ 78\right] & B &  B & B & A \\
\end{array}
\right), 
u_2 = \left(
\begin{array}{ccccc}
& \left[13 \right] & \left[24 \right] & \left[57 \right] & 
\left[68 \right] \nonu \\
 \left[13 \right] & C & -D & -i E &  -iE \nonu \\
\left[24 \right]  & -D & C & iE & iE \nonu \\
 \left[ 57\right] & iE & -iE & C & D \nonu \\
 \left[ 68\right] & iE &  -iE & D & C \\
\end{array}
\right), \nonu \\
&& u_3 = \left(
\begin{array}{ccccc}
& \left[14 \right] & \left[23 \right] & \left[58 \right] & 
\left[67 \right] \nonu \\
 \left[14 \right] & C & D & -iE &  iE \nonu \\
\left[23 \right]  & D & C & -iE & iE \nonu \\
 \left[ 58 \right] & iE & iE & C & -D \nonu \\
 \left[ 67 \right] & -iE & -iE & -D & C \\
\end{array}
\right), 
u_4 = \left(
\begin{array}{ccccc}
& \left[15 \right] & \left[26 \right] & \left[37 \right] & 
\left[48 \right] \nonu \\
 \left[15 \right] & C & -D & iE &  iE \nonu \\
\left[26 \right]  & -D & C & -iE & -iE \nonu \\
 \left[37\right] & -iE & iE & C & D \nonu \\
 \left[48 \right] & -iE &  iE & D & C \\
\end{array}
\right), \nonu \\
&& u_5 = \left(
\begin{array}{ccccc}
& \left[16 \right] & \left[25 \right] & \left[38 \right] & 
\left[47 \right] \nonu \\
 \left[16 \right] & C & D & iE &  -iE \nonu \\
\left[25 \right]  & D & C & iE & -iE \nonu \\
 \left[38 \right] & -iE & -iE & C & -D \nonu \\
 \left[47 \right] & iE & iE & -D  & C \\
\end{array}
\right), 
u_6 = \left(
\begin{array}{ccccc}
& \left[17 \right] & \left[28 \right] & \left[35 
\right] & \left[46 \right] \nonu \\
 \left[17 \right] & C & D & iE &  -iE \nonu \\
\left[28 \right]  & D & C & iE & -iE \nonu \\
 \left[35 \right] & -iE & -iE & C & -D \nonu \\
 \left[46 \right] & iE & iE & -D & C \\
\end{array}
\right), \nonu \\
&& u_7 = \left(
\begin{array}{ccccc}
& \left[18 \right] & \left[27 \right] & \left[36 \right] & 
\left[45 \right] \nonu \\
 \left[18 \right] & C & -D & iE &  iE \nonu \\
\left[27 \right]  & -D & C & -iE & -iE \nonu \\
 \left[36 \right] & -iE & iE & C & D \nonu \\
 \left[45 \right] & -iE &  iE & D & C 
\end{array} \right),
 v_1 = \left(
\begin{array}{ccccc}
& \left[12 \right] & \left[34 \right] & \left[56 \right] & 
\left[78 \right] \nonu \\
 \left[12 \right] & F & G & G &  G \nonu \\
\left[34 \right]  & G & F & G & 
G \nonu \\
 \left[ 56\right] & G & G & F & 
G \nonu \\
 \left[ 78\right] & G &  G & G & F \\
\end{array}
\right), \nonu \\ 
& & v_2 = \left(
\begin{array}{ccccc}
& \left[13 \right] & \left[24 \right] & \left[57 \right] & 
\left[68 \right] \nonu \\
 \left[13 \right] & H & -I & i J &  iJ \nonu \\
\left[24 \right]  & -I & H & -iJ & -iJ \nonu \\
 \left[ 57\right] & iJ & -iJ & -H & -I \nonu \\
 \left[ 68\right] & iJ &  -iJ & -I & -H \\
\end{array}
\right),  v_3 = \left(
\begin{array}{ccccc}
& \left[14 \right] & \left[23 \right] & \left[58 \right] & 
\left[67 \right] \nonu \\
 \left[14 \right] & H & I & iJ &  -iJ \nonu \\
\left[23 \right]  & I & H & iJ & -iJ \nonu \\
 \left[ 58 \right] & iJ & iJ & -H & I \nonu \\
 \left[ 67 \right] & -iJ & -iJ & I & -H \\
\end{array}
\right), \nonu \\
& & v_4 = \left(
\begin{array}{ccccc}
& \left[15 \right] & \left[26 \right] & \left[37 \right] & 
\left[48 \right] \nonu \\
 \left[15 \right] & H & -I & -iJ &  -iJ \nonu \\
\left[26 \right]  & -I & H & iJ & iJ \nonu \\
 \left[37\right] & -iJ & iJ & -H & -I \nonu \\
 \left[48 \right] & -iJ &  iJ & -I & -H \\
\end{array}
\right),
v_5 = \left(
\begin{array}{ccccc}
& \left[16 \right] & \left[25 \right] & \left[38 \right] & 
\left[47 \right] \nonu \\
 \left[16 \right] & H & I & -iJ &  iJ \nonu \\
\left[25 \right]  & I & H & -iJ & iJ \nonu \\
 \left[38 \right] & -iJ & -iJ & -H & I \nonu \\
 \left[47 \right] & iJ & iJ & I  & -H \\
\end{array}
\right), \nonu \\
& & v_6 = \left(
\begin{array}{ccccc}
& \left[17 \right] & \left[28 \right] & \left[35 
\right] & \left[46 \right] \nonu \\
 \left[17 \right] & -H & -I & iJ &  -iJ \nonu \\
\left[28 \right]  & -I & -H & iJ & -iJ \nonu \\
 \left[35 \right] & iJ & iJ & H & -I \nonu \\
 \left[46 \right] & -iJ & -iJ & -I & H \\
\end{array}
\right),  v_7 = \left(
\begin{array}{ccccc}
& \left[18 \right] & \left[27 \right] & \left[36 \right] & 
\left[45 \right] \nonu \\
 \left[18 \right] & -H & I & iJ &  iJ \nonu \\
\left[27 \right]  & I & -H & -iJ & -iJ \nonu \\
 \left[36 \right] & iJ & -iJ & H & I \nonu \\
 \left[45 \right] & iJ &  -iJ & I & H \\ 
\end{array} \right), \\
\label{uv}
\eea
where
\bea
& & A \equiv p^3, \;\;\; B \equiv pq^2, \;\;\; C \equiv pp'^2, \;\;\;
D \equiv pq'^2, \;\;\;E \equiv qp'q'
\nonu \\
& & F \equiv q^3, \;\;\;G \equiv p^2q, \;\;\;H \equiv qq'^2, \;\;\;
I \equiv qp'^2, \;\;\;J \equiv pp'q' \nonu
\eea
and $p, p', q$ and $q'$ are given in (\ref{sinhcosh}).
The nonzero components of $A_2$ tensor, $A_{2, L}^{\;\;\;\;\;IJK}$ 
can be obtained from
(\ref{ttensor}) and they are given by
\bea
&& A_{2, 1}^{\;\;\;\;\;\;256}=
A_{2, 1}^{\;\;\;\;\;234}=
A_{2, 2}^{\;\;\;\;\;\;165}=
A_{2, 2}^{\;\;\;\;\;\;143}=A_{2, 3}^{\;\;\;\;\;456}=
A_{2, 3}^{\;\;\;\;\;\;412}=
A_{2, 4}^{\;\;\;\;\;365}=A_{2, 4}^{\;\;\;\;\;321}=
\nonu \\
&& A_{2, 5}^{\;\;\;\;\;\;634}=
A_{2, 5}^{\;\;\;\;\;612}=
A_{2, 6}^{\;\;\;\;\;\;543}=A_{2, 6}^{\;\;\;\;\;521}
\equiv  y_1, \nonu \\
&& A_{2, 7}^{\;\;\;\;\;\;182}=
A_{2, 7}^{\;\;\;\;\;384}=A_{2, 7}^{\;\;\;\;\;586}=
A_{2, 8}^{\;\;\;\;\;127}=
A_{2, 8}^{\;\;\;\;\;\;473}=
A_{2, 8}^{\;\;\;\;\;675}
\equiv y_2, \nonu \\
&& A_{2, 1}^{\;\;\;\;\;\;728}=
A_{2, 2}^{\;\;\;\;\;\;817}=
A_{2, 3}^{\;\;\;\;\;\;748}=
A_{2, 4}^{\;\;\;\;\;837}=
A_{2, 5}^{\;\;\;\;\;768}=
A_{2, 6}^{\;\;\;\;\;857}
\equiv y_3, \nonu \\
&& A_{2, 1}^{\;\;\;\;\;\;368}=A_{2, 1}^{\;\;\;\;\;458}=
A_{2, 1}^{\;\;\;\;\;357}=
A_{2, 1}^{\;\;\;\;\;\;647}=
A_{2, 2}^{\;\;\;\;\;864}=
A_{2, 2}^{\;\;\;\;\;637}=
A_{2, 2}^{\;\;\;\;\;\;547}=A_{2, 2}^{\;\;\;\;\;358}=
\nonu \\
&& A_{2, 3}^{\;\;\;\;\;\;861}=
A_{2, 3}^{\;\;\;\;\;726}=
A_{2, 3}^{\;\;\;\;\;\;175}=A_{2, 3}^{\;\;\;\;\;285}=
A_{2, 4}^{\;\;\;\;\;682}=
A_{2, 4}^{\;\;\;\;\;185}=
A_{2, 4}^{\;\;\;\;\;\;725}=A_{2, 4}^{\;\;\;\;\;716}=
\nonu \\
&& A_{2, 5}^{\;\;\;\;\;\;427}=A_{2, 5}^{\;\;\;\;\;814}=
A_{2, 5}^{\;\;\;\;\;\;713}=A_{2, 5}^{\;\;\;\;\;823}=
A_{2, 6}^{\;\;\;\;\;428}=
A_{2, 6}^{\;\;\;\;\;813}=
A_{2, 6}^{\;\;\;\;\;\;273}=A_{2, 6}^{\;\;\;\;\;174}
\equiv y_4,  \nonu \\
&& A_{2, 7}^{\;\;\;\;\;\;362}=A_{2, 7}^{\;\;\;\;\;452}=
A_{2, 7}^{\;\;\;\;\;531}=
A_{2, 7}^{\;\;\;\;\;\;146}=A_{2, 8}^{\;\;\;\;\;163}=
A_{2, 8}^{\;\;\;\;\;541}=
A_{2, 8}^{\;\;\;\;\;\;532}=A_{2, 8}^{\;\;\;\;\;462}
\equiv y_5, \nonu  
\eea 
where
$y_i$'s are given in (\ref{yis}).
Notice that we did not write down other components of $A_2$ tensor 
which are interchanged
between 2nd and 3rd indices because it is manifest that 
$A_{2, L}^{\;\;\;\;\;IJK}=-
A_{2, L}^{\;\;\;\;\;IKJ}$, 
by definition. Moreover there is a symmetry between the upper indices:
$A_{2, L}^{\;\;\;\;\;IJK}=A_{2, L}^{\;\;\;\;\;JKI}=A_{2, L}^{\;\;\;\;\;KIJ}$. 

The kinetic term can be summarized as following block diagonal matrices:
\bea
A_{\mu}^{\;\;\;IJKL} = \mbox{diag} \left(A_{\mu,1}, A_{\mu,2},
A_{\mu,3}, A_{\mu,4}, A_{\mu,5}, A_{\mu,6}, A_{\mu,7} \right),
\nonu
\eea
where
\bea
&& A_{\mu,1} =\frac{1}{2} \partial_{\mu} \left(
\begin{array}{cccccc}
 & \left[12 \right] & \left[34 \right] & \left[56 \right] & 
\left[78 \right] \nonu \\  
\left[12 \right] & 0 & -  \la &   - \la &   -\la \nonu \\
\left[34 \right]  &  -\la & 0 &  
  -\la & -\la \nonu \\
\left[56 \right] &   -\la &  -\la 
& 0 & 
 - \la \nonu \\
\left[78 \right] &  -\la &   -
\la &  - \la & 0 \\
\end{array}
\right), 
A_{\mu,2} = \frac{1}{2} \partial_{\mu} \left(
\begin{array}{cccccc}
 & \left[13 \right] & \left[24 \right] & \left[57 \right] & 
\left[68 \right]  \nonu \\
\left[13 \right] & 0 &  \la  & -i \la' &   - i 
 \la' \nonu \\
\left[24 \right] &   \la  & 0 &  i 
 \la' &   i  \la' \nonu \\
\left[57 \right] &  -i  \la'  & i 
 \la' & 0 &   \la \nonu \\
 \left[68 \right] & -i  \la' &  i 
 \la' &   \la & 0 \\
\end{array}
\right), \nonu \\
&& A_{\mu,3} =\frac{1}{2} \partial_{\mu} \left(
\begin{array}{cccccc}
& \left[14 \right] & \left[23 \right] & \left[58 \right] &  \left[67 
\right] \nonu \\
\left[14 \right] & 0 &- \la  & -i  \la' & i 
\la' \nonu \\
\left[23 \right] & - \la  & 0 & -i  \la' & i  \la' \nonu \\
\left[58 \right] &  -i  \la' & -i \la' & 0 & -\la \nonu \\
\left[67 \right] & i \la' & i 
\la' & -\la & 0 \\
\end{array}
\right), 
A_{\mu,4} = \frac{1}{2} \partial_{\mu} \left(
\begin{array}{cccccc}
& \left[15 \right] & \left[26 \right] & \left[37 \right] & \left[48 
\right] \nonu \\
\left[15 \right] &  0 & \la & i  \la' & i \la' \nonu \\
\left[26 \right] &  \la & 0 & -i  \la' & -i \la' \nonu \\
\left[37 \right] & i \la'  & -i \la' & 0 & \la \nonu \\
\left[48 \right] &  i \la'  &  -i 
\la' &  \la & 0 \\
\end{array}
\right), \nonu \\
&& A_{\mu,5} =\frac{1}{2} \partial_{\mu}  \left(
\begin{array}{cccccc}
& \left[16 \right] & \left[25 \right] & \left[38 \right] & 
\left[47 \right] \nonu \\
\left[16 \right] &  0 &  -\la & i \la'  &  -i 
\la' \nonu \\
\left[25 \right] & -\la & 0 & i 
\la'  & -i \la' \nonu \\
\left[38 \right] & i  \la'  & i 
 \la'  & 0 & - \la \nonu \\
 \left[47 \right] & -i 
\la' & -i  \la' & - \la  & 0 \\
\end{array}
\right), 
A_{\mu,6} = -\frac{1}{2} \partial_{\mu}  \left(
\begin{array}{cccccc}
& \left[17 \right] & \left[28 \right] & \left[35 \right] & 
\left[46 \right] \nonu \\
\left[17 \right] &  0 &  -\la & i \la'  &  -i 
\la' \nonu \\
\left[28 \right] & -\la & 0 & i 
\la'  & -i \la' \nonu \\
\left[35 \right] & i  \la'  & i 
 \la'  & 0 & - \la \nonu \\
 \left[46 \right] & -i 
\la' & -i  \la' & - \la  & 0 \\
\end{array}
\right), \nonu \\
&& A_{\mu,7} =- \frac{1}{2} \partial_{\mu} \left(
\begin{array}{cccccc}
& \left[18 \right] & \left[27 \right] & \left[36 \right] & \left[45 
\right] \nonu \\
\left[18 \right] &  0 & \la & i  \la' & i \la' \nonu \\
\left[27 \right] &  \la & 0 & -i  \la' & -i \la' \nonu \\
\left[36 \right] & i \la'  & -i \la' & 0 & \la \nonu \\
\left[45 \right] &  i \la'  &  -i 
\la' &  \la & 0 
\end{array}
\right).
\eea

\vspace{2cm}
\centerline{\bf Acknowledgments} 

This work was supported 
by Korea Research Foundation Grant(KRF-1999-015-DI0019).
CA thanks Yonsei Visiting Research Center(YVRC) where this work was initiated.
We thank S.-J. Rey for helpful discussions on relating subjects and 
reading an earlier version of this  manuscript. 
We are grateful to H. Nicolai for helpful 
correspondence on his work.

\end{document}